\documentclass{article}
\usepackage{iclr2026_conference,times}

\usepackage[table,dvipsnames]{xcolor}
\definecolor{INK}{HTML}{1B2A32}
\definecolor{DECODE}{HTML}{0E7C7B}
\definecolor{DEPLOY}{HTML}{C44E52}
\definecolor{INERT}{HTML}{8A9AA5}
\definecolor{CONFOUND}{HTML}{E0A32E}
\definecolor{SCALE}{HTML}{6A4C93}
\definecolor{GRIDC}{HTML}{D8DEE0}
\definecolor{INKSOFT}{HTML}{4A5A63}
\definecolor{DECODEbg}{HTML}{E4F0F0}
\definecolor{DEPLOYbg}{HTML}{F7E3E4}
\definecolor{CONFOUNDbg}{HTML}{FBF1DC}
\definecolor{SCALEbg}{HTML}{EBE5F1}
\definecolor{HEADbg}{HTML}{EEF2F3}
\newcommand{\present}[1]{\textcolor{DECODE}{#1}}
\newcommand{\deployed}[1]{\textcolor{DEPLOY}{#1}}

\newcommand{\confound}[1]{\textcolor{CONFOUND}{#1}}
\newcommand{\atscale}[1]{\textcolor{SCALE}{#1}}

\usepackage{amsmath,amssymb,amsthm}
\usepackage{booktabs}
\usepackage{makecell}
\usepackage{multirow}
\usepackage{pifont}     
\usepackage{graphicx}
\usepackage{url}
\usepackage{xspace}
\usepackage{enumitem}
\usepackage{algorithm}
\usepackage{algpseudocode}
\usepackage{microtype}
\usepackage[hidelinks]{hyperref}
\setcitestyle{numbers,square,comma,sort&compress}

\DeclareMathOperator{\corr}{corr}
\newcommand{\R}{\mathbb{R}}
\newcommand{\eqdef}{\triangleq}
\newcommand{\uhat}{\hat{u}_{\ell}}
\newcommand{\hl}{h_{\ell}}
\newcommand{\dg}{\Delta g}
\newcommand{\dphi}{\Delta\phi}
\newcommand{\Rsq}{R^{2}}
\newcommand{\tcs}{\ensuremath{\tau}}                 
\newcommand{\phivar}{\phi}
\newcommand{\Poseidon}{Poseidon-B\xspace}

\iclrfinalcopy

\title{The Objective Decides:\\ When a Learned Dynamics Model Uses a Conserved Quantity}

\author{Chih-Ting Liao \& Xin Cao\\
University of New South Wales\\
\texttt{mill.liao@unsw.edu.au}}

\begin{document}
\maketitle
\lhead{Preprint}

\begin{abstract}
A linear probe that recovers a conserved quantity from a learned dynamics model's activations is
routinely read as evidence that the model \emph{uses} that quantity. We show this inference is
unsound. Across mechanical, circuit, and partial-differential-equation (PDE) systems, and on a
$158$M-parameter pretrained PDE foundation model, energy and other invariants are linearly decodable
at $\Rsq\!\approx\!1$ yet causally \present{inert} on next-state prediction: overwriting the decoded
direction with a donor state's value, single-step activation interchange, leaves the forward pass
essentially unchanged (transfer-corr $\tcs\!\approx\!0$). The \emph{same direction in the same
representation} becomes causally \deployed{load-bearing} ($\tcs\!\to\!+1$) the moment the training
objective rewards the invariant, so deployment is a property of the \emph{objective}, not of the
representation or the probe. We further show that \emph{when} an invariant is deployed is governed by a
precise algebraic predicate, its relation to the prediction output, by flipping a single invariant
from inert to load-bearing by changing only the output's algebra. Finally, the gap has teeth: across
models that all decode the target at $\Rsq\!=\!1.00$, the deployment gap forecasts out-of-distribution
(OOD) accuracy ($r\!=\!+0.97$) where decodability is blind. We argue that causal \emph{deployment}, not
decodability, is what interpretability should measure when the question is whether a model uses a piece
of knowledge, and we give a cheap instrument for measuring it.
\end{abstract}

\section{Introduction}
Does a neural network that predicts the physical world \emph{understand} the quantities that govern it?
The question is not academic. Learned dynamics models and PDE foundation models are increasingly used as
fast surrogates for simulation in the physical sciences, and a growing literature asks whether they have
internalized the conserved quantities, energy, momentum, charge, that a physicist would insist on. When
a probe recovers such a quantity from a model's activations, the field reads it as a yes: the model has
learned, represents, and uses the invariant \citep{alain2017probes,belinkov2022probing,iten2020scinet}.
That inference is what licenses trusting the surrogate. This paper argues that the inference is wrong, and
quantifies exactly how wrong.

The field's default answer comes from \emph{probing}: fit a linear map from a model's activations to a
quantity of interest and read a high $\Rsq$ as evidence that the model has learned and uses it. We show
this reading conflates two claims that come apart, and give an instrument that separates them.

\noindent\textbf{Move 1: presence is not use.}
A probe that recovers $\phivar$ at $\Rsq\!\approx\!1$ certifies that the information is \emph{present}
in the representation. It says nothing about whether the computation \emph{depends} on it. These are
different claims, and only the second predicts behavior. We make the distinction operational with a
causal test, single-step activation interchange, and find that in a fully-observed learned dynamics
model, energy is linearly decodable at $\Rsq\!\approx\!1$ and yet \emph{causally inert on next-state
prediction}: setting the decoded direction to any other in-distribution value changes the model's
one-step prediction by an amount statistically indistinguishable from a matched random direction. The
model represents the invariant and does not use it.

\noindent\textbf{Move 2: this is not a tautology.}
A skeptic will object that the dissociation is trivial, a next-state predictor is trained on the
microstate, so of course the invariant is redundant. Three facts make it non-trivial. \emph{(i)} The
\deployed{same direction} in the \deployed{same representation} flips from inert to causally
load-bearing when only the training objective changes, with an identical probe and identical evaluation
data; the effect is therefore a property of \emph{deployment}, not of the representation or the probe.
\emph{(ii)} Whether an invariant is deployed is governed by a precise, testable rule, its algebraic
relation to the prediction output, and we flip a \emph{single} invariant across this boundary by
changing the output's algebra alone. \emph{(iii)} The gap has downstream teeth: two models that decode
the correct feature at $\Rsq\!=\!1.00$ differ by $7.2\times$ in OOD accuracy, and only the deployment
metric separates them. A tautology has no predictive consequences; this does.

\noindent\textbf{Move 3: measure deployment, not decodability.}
We therefore argue that \textbf{causal deployment, not decodability, is the quantity interpretability
should measure} when the question is whether a model uses a piece of knowledge. Decodability is a test
of the representation; deployment is a test of the computation. Our contributions are:
\noindent\textbf{Contributions.} We contribute (i)~a conceptual and empirical separation of \emph{decodability} from \emph{deployment}, and a cheap single-step instrument to measure the latter: donor-patch activation interchange, read out with a scale-free transfer-corr $\tcs$ and calibrated against a matched random-direction control (\S\ref{sec:prelim}, \S\ref{sec:method}); (ii)~the decode--deploy \emph{dissociation} across substrates (mechanics, circuits, PDEs), dimensions ($2\!\to\!32$), invariant classes (energy, angular momentum), and architectures (MLP, GRU, Transformer), with every system decodable at $\Rsq\!\ge\!0.8$ yet inert on next-state prediction and load-bearing under an invariant-relevant objective (Table~\ref{tab:main}); (iii)~an \emph{algebraic taxonomy} of when an invariant is deployed, governed by its relation to the output rather than its identity, flipping one invariant from inert to load-bearing by changing only the output's algebra (\S\ref{sec:taxonomy}); (iv)~replication on \Poseidon, a $158$M-parameter pretrained PDE foundation model, on \emph{real} Navier--Stokes trajectories, with a depth-resolved structure that recovers both halves of the dissociation (\S\ref{sec:scale}); and (v)~a downstream consequence, the deployment gap \emph{forecasts} OOD failure ($r\!=\!+0.97$) where decodability is blind ($\Rsq\!=\!1.00$ for every model), turning an interpretability quantity into a predictor of generalization (\S\ref{sec:payoff}).

\section{Related Work}\label{sec:related}
\noindent\textbf{Probing physics in learned models.}
Probing classifiers read latent structure from representations with supervised readouts
\citep{alain2017probes,conneau2018probing,tenney2019bert,hewitt2019structural,belinkov2022probing}, and
control tasks caution that decodability can reflect the probe rather than the model
\citep{hewitt2019control}. In physics models specifically, SciNet recovers physical parameters from a
designed bottleneck \citep{iten2020scinet}; Hamiltonian, Lagrangian, and Noether architectures build
conservation in by construction \citep{greydanus2019hnn,cranmer2020lnn,alet2021noether}; and
symbolic-regression and sparse-dynamics pipelines extract governing laws or conserved quantities from data
\citep{cranmer2020discovering,udrescu2020aifeynman,brunton2016sindy,liu2021aipoincare,ha2021conservnet,lemos2023orbital}.
Closest in spirit, \citet{vafa2024worldmodel,vafa2025inductivebias} probe whether a sequence model trained
on orbital mechanics carries the right inductive bias, finding accurate prediction without recovery of
Newtonian structure. All of this asks what is \emph{present} in the representation, or engineers presence
directly; none isolates whether a decodable invariant is \emph{causally deployed}, whether that depends on
the objective, or what governs it.

\noindent\textbf{Causal interpretability.}
Our instrument is standard: activation patching and interchange interventions
\citep{vig2020mediation,geiger2021causal,meng2022rome,wang2023ioi}, formalized as causal abstraction and
its learned variants \citep{geiger2023das,wu2023boundless}, within a broader circuit- and feature-level
program \citep{olah2020circuits,nanda2023grokking,conmy2023acdc,raukert2023survey}. Emergent-world-model
studies combine probing with intervention to argue a model uses a learned state, in board games, chess,
navigation, and programs \citep{li2023othello,nanda2023othello,toshniwal2022chess,jin2024program,gurnee2024space},
often finding linearly represented state \citep{park2023linear,mikolov2013word2vec}. We do \emph{not} claim
the intervention method as a contribution; we adopt it, add a scale-free readout and a matched control, and
apply it to a question these works do not ask, the objective- and algebra-dependence of deployment for a
\emph{continuous} conserved quantity, and its link to generalization.

\noindent\textbf{Superposition, sparse features, and shortcuts.}
That a decodable feature can be causally inert is consistent with distributed, redundant coding
\citep{elhage2022superposition,bricken2023monosemanticity,cunningham2023sparse,templeton2024scaling,gao2024scaling}
and with the observation that models solve tasks by shortcuts rather than the intended mechanism
\citep{geirhos2020shortcut}. We make this precise for physical invariants and, crucially, show the
deployment gap is \emph{predictive} of OOD behavior. Our delta over the closest work, objective-dependence,
an algebraic taxonomy of \emph{when} it holds, foundation-model replication, and a deployment-to-OOD link,
is summarized in Table~\ref{tab:related}.

\newcommand{\cyes}{\textcolor{DECODE}{\ding{51}}}
\newcommand{\cno}{\textcolor{INERT}{\ding{55}}}
\newcommand{\cpart}{\textcolor{CONFOUND}{$\sim$}}

\begin{table}[t]
\centering
\footnotesize
\setlength{\tabcolsep}{4pt}
\renewcommand{\arraystretch}{1.15}
\caption{\textbf{Where this work sits.} Prior work reads what is \emph{present} in a physics model's
representation, or builds conservation in by design; none isolates the \emph{objective-dependence of
causal deployment} of a decodable invariant, characterizes \emph{when} it is deployed, and shows the
deployment gap forecasts out-of-distribution (OOD) failure. \cyes\,=\,yes, \cpart\,=\,partial, \cno\,=\,no.}
\label{tab:related}
\begin{tabular}{@{}p{2.28in}ccccc@{}}
\toprule
 & \makecell{decode /\\present?} & \makecell{causal\\interv.} & \makecell{obj.-dep.\\deploy} & \makecell{algebr.\\taxon.} & \makecell{gap\\$\to$OOD} \\
\midrule
SciNet, HNN, Noether Nets \citep{iten2020scinet,greydanus2019hnn,alet2021noether} & \cyes & \cno & \cno & \cno & \cno \\
AI-Feynman, symbolic discovery \citep{cranmer2020discovering,udrescu2020aifeynman,liu2021aipoincare} & \cyes & \cno & \cno & \cno & \cno \\
World-model probes \citep{vafa2024worldmodel,vafa2025inductivebias} & \cyes & \cpart & \cno & \cno & \cpart \\
Invariant recovery \citep{vafa2025inductivebias,lemos2023orbital} & \cyes & \cno & \cno & \cno & \cno \\
Causal abstraction, DAS \citep{geiger2021causal,geiger2023das} & \cyes & \cyes & \cno & \cno & \cno \\
\rowcolor{DEPLOYbg} \textbf{This work} & \cyes & \cyes & \cyes & \cyes & \cyes \\
\bottomrule
\end{tabular}
\end{table}

\begin{table}[t]
\centering
\small
\setlength{\tabcolsep}{4.5pt}
\renewcommand{\arraystretch}{1.1}
\caption{\textbf{The decode--deploy dissociation across substrates, dimensions, invariant classes, and
architectures.} We report transfer-corr $\tau$ (scale-free; $\approx 0$ inert, $\to +1$ load-bearing).
Every system is decodable ($R^2\!\ge\!0.8$) yet \present{inert on next-state prediction} and
\deployed{load-bearing on an invariant-relevant objective}. \confound{Linear/extensive invariants} are a
separately analyzed confounded boundary (decode direction entangled with the linear output map). The
at-scale rows use Poseidon-B, a $158$M-parameter PDE foundation model, on real incompressible
Navier--Stokes trajectories. Verified numbers; per-run detail and $95\%$ CIs in App.~\ref{app:results}.}
\label{tab:main}
\begin{tabular}{llccc}
\toprule
System & Invariant (class) & decode $R^2$ & \makecell{next-state $\tau$\\(inert)} & \makecell{invariant obj.\ $\tau$\\(load-bearing)} \\
\midrule
pendulum        & energy (nonlinear)          & $1.00$ & \present{$-0.01$} & \deployed{$+0.97$} \\
LC circuit      & energy (nonlinear)          & $0.99$ & \present{$+0.00$} & \deployed{$+0.31^{\dagger}$} \\
central force   & angular mom.\ (bilinear)    & $0.99$ & \present{$+0.02$} & \deployed{$+1.00$} \\
wave PDE ($32$-d) & energy (quadratic)        & $0.99$ & \present{$+0.00$} & \deployed{$+0.97$} \\
Kepler $1/r^2$     & energy (entangled)          & $0.80$ & \present{$-0.02$} & \deployed{$+0.39$} \\
GRU / Transformer  & energy (history window)     & $0.98$ & \present{$+0.02$} & \deployed{$+0.67$} \\
\midrule
\rowcolor{CONFOUNDbg} spring / wave & lin.\ momentum \emph{(linear)} & $0.98$ & \confound{$-0.71$} & \deployed{$+1.00$} \\
\midrule
\multicolumn{5}{l}{\atscale{\textbf{Poseidon-B}~\citep{herde2024poseidon}:}} \\
\atscale{$\;\;$bottleneck} & KE (nonlinear) + $\int u$ (linear) & $0.92$ & \present{$+0.06$} & \deployed{$+0.85^{\ddagger}$} \\
\atscale{$\;\;$decoder (output-adj.)} & $\int u$ \emph{(linear)} & $1.00$ & \confound{$+0.36$} & $-$ \\
\bottomrule
\end{tabular}
\end{table}

\section{Problem Formulation: Decodable versus Deployed}\label{sec:prelim}
\noindent\textbf{Setup.}
A learned dynamics model $f_\theta$ maps a state $s_t\in\mathcal{S}\subseteq\R^{d}$ (or a fixed history
window) to a next-state prediction $\hat{s}_{t+1}=f_\theta(s_t)$. Writing $\hl(\cdot)\in\R^{m}$ for the
hidden activation at layer $\ell$, the model factors as $f_\theta=f^{>\ell}\!\circ\hl$. A conserved
quantity is a function $\phivar:\mathcal{S}\to\R$ (e.g.\ total energy) with
$\phivar(s_t)\approx\phivar(s_{t+1})$ along true trajectories. We want to know whether $f_\theta$'s
computation \emph{uses} $\phivar$.

\noindent\textbf{Decodability (presence).}
$\phivar$ is $\epsilon$-\emph{decodable} at layer $\ell$ if there exist $w_\ell\in\R^{m},b\in\R$ with
cross-validated
\begin{equation}
\Rsq\!\left(w_\ell^{\top}\hl(s)+b,\ \phivar(s)\right)\ \ge\ 1-\epsilon .
\label{eq:decode}
\end{equation}
The fitted \emph{invariant direction} is $\uhat=w_\ell/\lVert w_\ell\rVert$. Equation~\eqref{eq:decode}
is the quantity the probing literature reports. It is a statement about the representation only.

\noindent\textbf{Deployment (use).}
Deployment asks whether the forward pass \emph{depends} on the value along $\uhat$. Given a target state
$s$ and a donor state $s'$, form the single-step \emph{interchange} representation that overwrites the
component along $\uhat$ with the donor's,
\begin{equation}
\hl^{\,s\to s'}\ =\ \hl(s)\ +\ \uhat\Big(\langle\uhat,\hl(s')\rangle-\langle\uhat,\hl(s)\rangle\Big),
\label{eq:interchange}
\end{equation}
and complete the forward pass, $\hat{s}_{t+1}^{\,s\to s'}=f^{>\ell}\!\big(\hl^{\,s\to s'}\big)$. Let
$g(\cdot)\in\R$ be a scalar readout of the prediction, either the model's own next-state functional or
an invariant-relevant target (below). Regress the induced change in $g$ on the change in the donor's
invariant,
\begin{equation}
\dg=g\big(\hat{s}_{t+1}^{\,s\to s'}\big)-g\big(\hat{s}_{t+1}^{\,s}\big),
\qquad
\dphi=\phivar(s')-\phivar(s).
\end{equation}
We report the OLS \emph{interchange slope} $\beta=\partial\,\dg/\partial\,\dphi$ and, as the metric of
record, the scale-free \emph{transfer-corr}
\begin{equation}
\boxed{\ \tcs\ \eqdef\ \corr\!\big(\dg,\ \dphi\big)\ \in\ [-1,1].\ }
\label{eq:tc}
\end{equation}
A direction is \deployed{load-bearing} iff $\beta>0$ and $\tcs\to+1$ (the output tracks the donor's
invariant); it is \present{inert} iff $\beta\approx0$ and $\tcs\approx0$ (the forward pass ignores the
overwrite). Because $\tcs$ is a correlation it is invariant to the arbitrary scale of $\uhat$ and of the
readout, which matters when comparing across systems of different dimension.

\noindent\textbf{The decode--deploy gap.}
Let $\tcs^{\text{next}}_\ell$ use the model's own next-state readout and $\tcs^{\text{inv}}_\ell$ use an
invariant-relevant objective. The \emph{deployment gap} at layer $\ell$ is
\begin{equation}
\gamma_\ell\ \eqdef\ \tcs^{\text{inv}}_\ell-\tcs^{\text{next}}_\ell .
\label{eq:gap}
\end{equation}
The central phenomenon of this paper is that $\gamma_\ell$ is large while
Eq.~\eqref{eq:decode} holds ($\Rsq\!\approx\!1$) under \emph{both} readouts: the same, decodable
direction is inert for one objective and load-bearing for another.

\noindent\textbf{An algebraic predicate for deployment.}
Whether an invariant is deployed on next-state prediction is not arbitrary. Let $y=g\text{-coords}(\hat
s_{t+1})$ be the output coordinates the readout sees. Call $\phivar$ \emph{output-entangled} if it is
(approximately) affine in those coordinates on-distribution, $\phivar(s)\approx a^{\top}y+c$; then
interchanging $\uhat$ mechanically moves $g$ and $\tcs$ is nonzero for a purely algebraic reason (a
\emph{confound}, and possibly negative from sign). Call it \emph{output-separated} if $\phivar$ is a
genuinely nonlinear function of $y$ with a vanishing linear part on-distribution; then next-state
interchange is \present{inert}. This predicate, not the identity of the invariant, predicts the
next-state verdict (\S\ref{sec:taxonomy}), and it is orthogonal to deployment under an invariant-relevant
objective, which reflects genuine use.

\section{Method: A Single-Step Deployment Probe}\label{sec:method}
\begin{figure}[t]
\centering
\includegraphics[width=\textwidth]{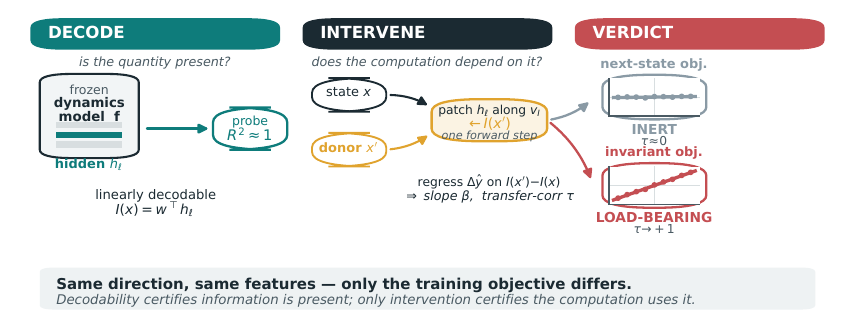}
\caption{\textbf{The instrument} (formalized in Algorithm~\ref{alg:probe}). \emph{Decode} fits a linear
invariant direction $\uhat$; \emph{intervene} overwrites its component with a donor's value and runs one
forward step; \emph{verdict} regresses the readout change on the donor's invariant change, flat,
$\tcs\!\approx\!0$ means \present{present but inert}; positive, $\tcs\!\to\!+1$ means \deployed{causally
load-bearing}.}
\label{fig:schematic}
\end{figure}

The instrument is summarized in Figure~\ref{fig:schematic} and given in full as Algorithm~\ref{alg:probe};
it has three parts, decode, intervene, verdict, and two controls, all on frozen weights with one forward
step per interchange. We fit the decode direction by $z$-scored
cross-validated ridge, which also locates the layer where $\phivar$ is most linearly present, then apply
the single-step interchange over many (target, donor) pairs for two readouts, the model's own next-state
output and a head trained to output $\phivar$ from the frozen features, summarizing each by the scale-free
$\tcs$ (Eq.~\ref{eq:tc}) so that systems of different dimension are comparable.

\begin{algorithm}[t]
\caption{Single-Step Deployment Probe (decodability vs.\ causal use of an invariant $\phivar$)}
\label{alg:probe}
\begin{algorithmic}[1]
\Require frozen $f_\theta\!=\!f^{>\ell}\!\circ\hl$; layer $\ell$; invariant $\phivar$; data $\mathcal{D}$; readouts $g_{\text{next}}$ (model's own output), $g_{\text{inv}}$ (head trained to output $\phivar$)
\State $w_\ell \gets z$-scored CV ridge of $\phivar(s)$ on $\hl(s)$;\ \ $\uhat\gets w_\ell/\lVert w_\ell\rVert$;\ \ report held-out $\Rsq$ \Comment{presence, Eq.~\ref{eq:decode}}
\For{readout $g\in\{g_{\text{next}},\,g_{\text{inv}}\}$}
  \For{many $(s,s')$ pairs from $\mathcal{D}$}
    \State $\hl^{\,s\to s'}\gets \hl(s)+\uhat\big(\langle\uhat,\hl(s')\rangle-\langle\uhat,\hl(s)\rangle\big)$ \Comment{single-step interchange, Eq.~\ref{eq:interchange}}
    \State $\dg\gets g\!\big(f^{>\ell}(\hl^{\,s\to s'})\big)-g\!\big(f^{>\ell}(\hl(s))\big)$;\ \ $\dphi\gets\phivar(s')-\phivar(s)$
  \EndFor
  \State $\tcs_g\gets\corr(\dg,\dphi)$;\ \ $\beta_g\gets$ OLS slope $\partial\dg/\partial\dphi$
\EndFor
\State \textbf{Controls:} repeat with (i) $\uhat$ refit on shuffled $\phivar$ labels and (ii) a matched random unit direction; require $\tcs\!\approx\!0$ for both
\State \textbf{Verdict:} \present{inert} if $\beta_{\text{next}}\!\approx\!0,\ \tcs_{\text{next}}\!\approx\!0$;\ \ \deployed{load-bearing} if $\beta_{\text{inv}}\!>\!0,\ \tcs_{\text{inv}}\!\to\!+1$
\State \Return $\Rsq$,\ $(\tcs_{\text{next}},\tcs_{\text{inv}})$,\ deployment gap $\gamma=\tcs_{\text{inv}}-\tcs_{\text{next}}$ (Eq.~\ref{eq:gap})
\end{algorithmic}
\end{algorithm}

\noindent\textbf{Design rationale: why a single forward step.}
The interchange is deliberately one step, not a rollout. A single step isolates the \emph{direct} causal
effect of the invariant, without entangling the estimate with error compounding, the stability of the
learned flow, or how the optimizer shaped multi-step behavior, all of which a rollout would conflate, and
it makes the null exactly matched (a random unit direction of the same norm, patched in the same place, for
the same one step). The reading is then clean: \emph{does one step of the model's own computation carry the
donor's invariant into the output?} If not, the invariant is present but epiphenomenal. The probe is
correspondingly cheap, one forward step per pair on a frozen model, so it scales to a $158$M-parameter
foundation model without training (\S\ref{sec:scale}).

\noindent\textbf{Controls.}
Both null conditions in Algorithm~\ref{alg:probe}, a shuffled-label probe (refit on permuted $\phivar$)
and a matched random-direction interchange, must read $\approx0$ or the verdict is void; the
random-direction null also calibrates how much a generic edit of matched norm leaks into the output.

\noindent\textbf{Our estimates are conservative lower bounds.}
When $\phivar$ is redundantly encoded across many directions (\S\ref{sec:robust}), overwriting a
\emph{single} direction leaves parallel copies intact, so single-step interchange \emph{under-reads}
deployment: \textbf{our load-bearing $\tcs$ values are lower bounds}. The random-direction null also
inflates for $k\!>\!1$ interchanged directions (a redundant invariant leaks into random subspaces), which
is why the directed $\tcs$, not a ratio-to-random, is our metric of record.

\section{Experiments}\label{sec:exp}
We first establish the dissociation on a controlled anchor, then show it generalizes,
characterize \emph{when} it happens, replicate it at scale on a
pretrained foundation model, demonstrate its downstream payoff,
and close with robustness. Every number below is drawn from a verified experiment
log; per-run detail, seeds, and $95\%$ confidence intervals are in Appendix~\ref{app:results}.

\begin{figure}[t]
\centering
\includegraphics[width=\textwidth]{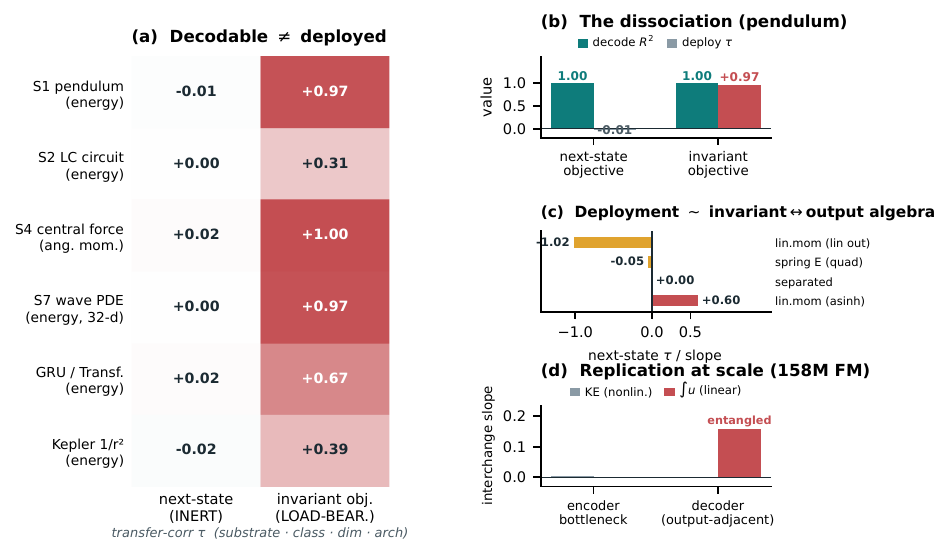}
\caption{\textbf{The decode--deploy dissociation, at a glance.}
\textbf{(a)}~Across six systems, every invariant is decodable yet \present{inert on next-state}
(left column, $\tcs\!\approx\!0$) and \deployed{load-bearing under an invariant-relevant objective}
(right column, $\tcs\!\to\!+1$). \textbf{(b)}~The pendulum anchor (\S\ref{sec:dissoc}): decode $\Rsq\!=\!1$
under both readouts; next-state inert, invariant-target load-bearing. \textbf{(c)}~The algebraic taxonomy
(\S\ref{sec:taxonomy}): the next-state verdict tracks the invariant's relation to the output, and the
\emph{same} linear momentum flips from \confound{confounded} to \deployed{load-bearing} when the output's
algebra changes. \textbf{(d)}~At scale on Poseidon-B~\citep{herde2024poseidon} ($158$M, \S\ref{sec:scale}): at the encoder
bottleneck both invariants are inert; at an output-adjacent decoder layer the \confound{linear} invariant
becomes entangled while the \present{nonlinear} one stays inert, the taxonomy, depth-resolved.}
\label{fig:results}
\end{figure}

\noindent\textbf{The dissociation, isolated.}\label{sec:dissoc}
Our anchor is an MLP trained to predict the next state of a nonlinear pendulum from its full state.
Energy is linearly decodable from the hidden layer at $\Rsq\!=\!1.00$. Yet on next-state prediction the
energy direction is \present{inert}: the interchange slope is $+0.0000$ and $\tcs\!=\!-0.01$, i.e.\ $0.3\times$
the matched random-direction null. Both controls read $\approx0$. Re-reading out the \emph{same frozen
features} against an invariant-relevant target (here the turning point $\theta_{\max}$, an
energy-equivalent) flips the verdict to \deployed{load-bearing}: slope $+0.0891$, $25.1\times$ random,
$\tcs\!=\!+0.97$. The model represents energy at $\Rsq\!=\!1$ and does not use it to predict the next
state; the identical representation is causally necessary once the objective rewards the invariant
(Figure~\ref{fig:results}b). This is the deployment gap of Eq.~\eqref{eq:gap} on a single system, with
the representation and probe held fixed.

\noindent\textbf{Why this is not a re-parameterization.} A natural worry is that flipping the readout target
merely relabels the same fact, that \emph{inert} and \emph{load-bearing} are two descriptions of one
representation. They are not. The decode fit, the hidden features, and the intervention are byte-for-byte
identical across the two verdicts; only the quantity we regress the post-step change onto differs. If the
verdict were a property of the representation, it could not move while the representation is frozen. What
moves it is the \emph{objective}: the pendulum's next-state map does not need energy (it is a smooth
function of the raw state that happens to conserve energy), whereas a map rewarded for the turning point
must route information through the invariant. The taxonomy of \S\ref{sec:taxonomy} makes this sharper
still, there, the \emph{same} conserved quantity flips verdict when we change only the algebra of the
output, holding both the representation and the objective's physical content fixed. Deployment is thus a
joint property of representation, objective, and output algebra, not a restatement of decodability.

\noindent\textbf{It generalizes across substrate, dimension, class, and architecture.}\label{sec:gen}

The dissociation is not an artifact of one system. We repeat the two-readout protocol across a mechanical
oscillator, an LC circuit, a $2$D central-force problem, a $32$-dimensional wave PDE, a Kepler $1/r^2$
system, and recurrent/attention architectures. Table~\ref{tab:main} and Figure~\ref{fig:results}a report
the result: decode $\Rsq\!\ge\!0.8$ everywhere, next-state $\tcs\!\approx\!0$ (inert) everywhere, and
invariant-objective $\tcs$ strongly positive (load-bearing) everywhere. Coverage spans \emph{substrate}
(mechanics / circuit / PDE), \emph{dimension} ($2\!\to\!4\!\to\!32$), \emph{invariant class} (energy,
angular momentum), and \emph{architecture}. Inertness on next-state is architecture-agnostic (MLP, GRU,
and Transformer all read $\tcs\!\approx\!0$), while the load-bearing magnitude is architecture-graded
(MLP $+0.97>$ GRU $+0.79>$ Transformer $+0.70$), consistent with the deployment head extracting the
invariant more or less cleanly from each backbone.

\begin{figure}[t]
\centering
\includegraphics[width=0.76\textwidth]{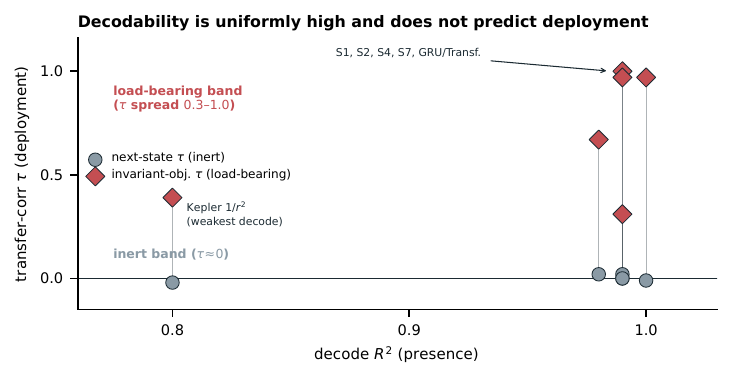}
\caption{\textbf{Decodability does not predict deployment.} Every system decodes its invariant at
$\Rsq\!\ge\!0.8$ (all points at high $x$), yet the \emph{same} high decodability is compatible with both a
\present{next-state-inert} reading ($\tcs\!\approx\!0$, lower band) and a \deployed{load-bearing} reading
($\tcs$ spread $0.3$--$1.0$, upper band). The vertical link for each system connects its two objectives:
presence is fixed and high; deployment is what varies. This is the decode--deploy gap of
Eq.~\eqref{eq:gap} across the full matrix.}
\label{fig:decouple}
\end{figure}

Presence itself tracks the objective: on the wave PDE the next-state network decodes energy at only
$\Rsq\!=\!0.13$, never building the feature it does not need, whereas an invariant-target network on the
\emph{same} substrate decodes it at $0.99$. Figure~\ref{fig:decouple} makes the decoupling explicit,
decodability is uniformly high and carries no information about whether the invariant is deployed.

\noindent\textbf{What each system controls for.}
The six systems are chosen so that a single explanation cannot cover them. The LC circuit shares the
pendulum's mathematics but not its physics, so agreement isolates a property of the \emph{objective}
rather than of the mechanical domain. The $2$D central-force problem swaps energy for angular momentum, a
\emph{bilinear} rather than quadratic invariant, testing a different algebraic class. The $32$-dimensional
wave PDE moves from particles to a spatially extended field, testing that the effect survives high
dimension. The Kepler $1/r^2$ system is chaotic and its energy is only partially decodable
($\Rsq\!=\!0.80$), yet the dissociation persists, showing it does not require a clean representation. The
GRU and Transformer must reconstruct the invariant across a history window rather than read it from a
single state, testing that inertness is not an artifact of the feed-forward inductive bias. Across all
six the pattern is identical: decodable, inert on next-state, load-bearing under the invariant objective.

\noindent\textbf{When is an invariant deployed? An algebraic taxonomy.}\label{sec:taxonomy}
Why is next-state prediction inert for energy but not for every invariant? The predicate of
\S\ref{sec:prelim} answers: the next-state verdict tracks the invariant's algebraic relation to the
output, and nothing else. Empirically this is a \emph{spectrum} of overlap, not a clean binary
(Figure~\ref{fig:results}c). Linear momentum, which is affine in the velocity outputs, produces a strong
\confound{confound} ($\tcs\!=\!-1.02$); spring energy, quadratic in those outputs, a weak one
($\tcs\!=\!-0.05$); invariants well-separated from the output (pendulum, LC, central force, wave) are
cleanly \present{inert} ($\approx0.00$).

The decisive test holds the invariant fixed and changes only the output's algebra. Taking the
\emph{same} linear momentum in the \emph{same} system and replacing the output map with a nonlinear
$\operatorname{asinh}(P)$ target flips the verdict from \confound{confounded} ($-1.02$) to
\deployed{load-bearing} ($+0.60$). Deployment on next-state prediction is therefore governed by the
invariant--output relation, not by the invariant's identity, a testable rule, not a vague appeal to
what the model \emph{needs.} We treat linear/extensive invariants as a separately analyzed
\confound{confounded boundary} (shaded row, Table~\ref{tab:main}) rather than folding them into the clean
cases.

This flip is the cleanest control in the paper. It holds the system, the state distribution, the model
class, and the invariant all fixed, and changes only the algebra of the quantity the model is asked to
output. Because everything a \emph{the model needs it} story would appeal to is unchanged, the swing from
$\tcs\!=\!-1.02$ to $\tcs\!=\!+0.60$ can only be attributed to deployment, the same information is
mechanically forced through the output in one case and genuinely recruited in the other. The overlap
spectrum and the flip together say that the next-state verdict is not measuring whether the invariant is
useful; it is measuring how much the invariant already leaks into the output coordinates, which is a
property we can compute and manipulate rather than an intrinsic fact about the physics.

\noindent\textbf{At scale: a 158M-parameter PDE foundation model on real trajectories.}\label{sec:scale}
To show the effect is not an artifact of models we train ourselves, we test Poseidon-B~\citep{herde2024poseidon}, a $158$M-parameter pretrained
PDE foundation model (an scOT SwinV2 backbone), on \emph{real} incompressible Navier--Stokes trajectories
(NS-Sines). Both halves of the dissociation replicate on real data: kinetic energy is decodable at
$\Rsq\!=\!0.92$ but causally \present{inert} on the model's own prediction ($\tcs\!=\!+0.06$), while an
invariant-relevant readout on the \emph{same frozen features} is \deployed{load-bearing}
($\tcs\!=\!+0.85$). Because the inputs are real trajectories, the result carries no synthetic-data
assumption; this is a pre-registered replication, not an emergent-capability claim.

The foundation model also exposes a \emph{depth} axis that the controlled systems cannot (Figure~\ref{fig:results}d). At
the encoder bottleneck, both a nonlinear invariant (KE, slope $+0.0032$) and a linear one ($\int u$, slope
$-0.0007$) are inert despite near-perfect decodability, the bottleneck stores the \emph{state}, not the
invariant. At an output-adjacent decoder layer, the \confound{linear} invariant becomes entangled with
the prediction (slope $+0.1582$, $\tcs\!=\!+0.36$, decode $\Rsq\!=\!1.00$) while the \present{nonlinear}
one remains inert (slope $+0.0002$). This is exactly the algebraic taxonomy of \S\ref{sec:taxonomy},
resolved along depth, in a model trained by others at scale.

The depth structure is itself informative: the bottleneck storing the state rather than the invariant is
what a good next-state predictor \emph{should} do (the microstate is a sufficient statistic, the invariant
redundant to it), so both invariants being inert there is expected, not surprising. A single frozen
checkpoint thus reproduces the dissociation, the objective-dependence, and the algebraic taxonomy at once,
on real data, with nothing fit but the linear probes.

\noindent\textbf{The payoff: deployment forecasts OOD failure where decodability is blind.}\label{sec:payoff}
\begin{figure}[t]
\centering
\includegraphics[width=\textwidth]{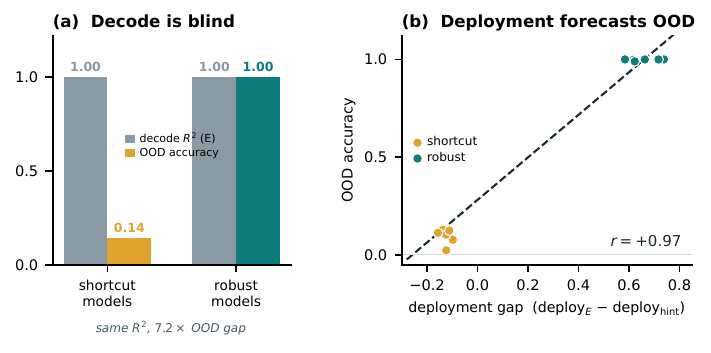}
\caption{\textbf{Deployment predicts generalization; decodability does not.}
\textbf{(a)}~Twelve models all decode the target invariant at $\Rsq\!=\!1.00$ (flat, left), so decode is
blind to a $7.2\times$ spread in OOD accuracy. \textbf{(b)}~The deployment gap forecasts OOD accuracy with
$r\!=\!+0.97$: \confound{shortcut} models (gap $\approx\!-0.13$) collapse OOD ($0.14$), while
\deployed{robust} models (gap $\approx\!+0.66$) generalize ($1.00$).}
\label{fig:payoff}
\end{figure}

Does the gap matter downstream? We take twelve models on a task with a spurious shortcut. Every one of
them decodes the task-relevant invariant at $\Rsq\!=\!1.00$, so decodability certifies all twelve
equally, and is \emph{blind} to their behavior. The \emph{deployment} gap is not: shortcut-reliant
models have a negative gap ($-0.13$) and collapse out of distribution (accuracy $0.14$), while models that
deploy the invariant have a positive gap ($+0.66$) and generalize (accuracy $1.00$), a $7.2\times$
difference. Across the twelve models the deployment gap correlates with OOD accuracy at $r\!=\!+0.97$
(Figure~\ref{fig:payoff}). An interpretability quantity that all-but-perfectly predicts generalization
is worth more than one that saturates at $\Rsq\!=\!1$ for everyone; a tautology could not do this.

\noindent\textbf{False assurance, made concrete.}
The practical stakes are sharp. A decode-based audit of these twelve models, the standard probing
protocol, returns $\Rsq\!=\!1.00$ for all of them and would certify every one as having \emph{learned} the
task-relevant invariant. Six of those twelve fail out of distribution. Decodability does not merely miss
the failures; it actively vouches for them. The deployment probe, applied to the same frozen models at no
additional training cost, separates the six that generalize from the six that do not. When a probe-based
claim of understanding is used to greenlight a scientific model, the quantity that should gate the
decision is deployment, not presence.

\noindent\textbf{Why the gap has predictive teeth.} The correlation is not a coincidence of this task; it has a
mechanistic reading. A model that solves the training distribution through a spurious shortcut routes its
prediction around the invariant, the invariant is present in its features (hence decodable) but off the
computational path (hence inert under interchange, giving a small or negative gap). Out of distribution,
the shortcut and the invariant come apart, and the shortcut-reliant model has nothing to fall back on. A
model that deploys the invariant routes its prediction \emph{through} it (large positive gap), so its
computation is anchored to a quantity that remains valid when the surface statistics shift. The
deployment gap thus measures, on frozen weights at training time, whether the model's computation is
organized around the invariant or merely stores it, which is why a quantity that saturates decode
$\Rsq$ for all twelve still separates them by $7.2\times$ downstream: decode reads the invariant's
presence, the gap reads its participation.

\noindent\textbf{Robustness: redundancy, statistics, and the linear boundary.}\label{sec:robust}
Three checks confirm the effect is real and correctly scoped; Figure~\ref{fig:redundancy} and
Appendix~\ref{app:robust} give details. \emph{Redundancy}: the deployed invariant is redundantly encoded
(self-repair index $0.76$--$0.999$; deployment strength non-monotonic in width, peaking at intermediate
capacity), which is why single-direction interchange under-reads and our load-bearing $\tcs$ are lower
bounds. A sparse-autoencoder latent aligns with the invariant at $|\text{corr}|\!=\!0.84$ with ablation
$7\times$ the random null, localizing the direction. \emph{Statistics}: a bootstrap and permutation pass
on the representative dissociation confirms the sign and significance structure with a deliberately
conservative single-direction estimator, inert conditions are statistically indistinguishable from zero
(e.g.\ $\tcs\!=\!-0.010$, $95\%$ CI $[-0.019,+0.001]$, $p\!\approx\!0.73$), while deployment conditions are
significantly positive ($p\!=\!0.0033$). \emph{Boundary}: linear/extensive invariants are entangled with
the linear output map ($\tcs\!=\!-0.71$ for spring/wave linear momentum) and we report them as a
confounded category, not as clean dissociations. \emph{Controls}: the inertness is not a probe artifact,
it holds across layers and $\{1,5,16\}$-dimensional subspaces, and a tight two-dimensional bottleneck
stores the state, not energy (decode $\Rsq\!\approx\!0.003$). The dissociation is clean for a structural
reason: next-state prediction on the microstate is the \emph{worst-case} objective for a load-bearing
invariant, since the microstate is a sufficient statistic that renders any conserved quantity redundant.

\section{Conclusion and Limitations}\label{sec:conclusion}
We have shown that a learned dynamics model can decode a conserved quantity at $\Rsq\!\approx\!1$ and not
use it; that the same, decodable direction becomes causally load-bearing when the objective rewards the
invariant; that \emph{when} an invariant is deployed is governed by a precise algebraic predicate we can
flip on demand; that both halves of the dissociation replicate on a $158$M-parameter pretrained PDE
foundation model on real trajectories; and that the deployment gap forecasts OOD failure where
decodability is blind. The practical message for interpretability is a change of measurement: when the
question is whether a model \emph{uses} a piece of knowledge, report causal deployment, not decodability.

\noindent\textbf{A recipe for practice.}
The deployment probe is a drop-in addition to any probing study. Given a trained model, a candidate
quantity $\phivar$, and the model's objective: fit the decode direction $\uhat$ as usual; then, at the
same layer, run the single-step interchange of Eq.~\eqref{eq:interchange} against the model's own output
and against an invariant-relevant readout, and report the two transfer-corrs and their gap
(Eq.~\ref{eq:gap}) alongside the decode $\Rsq$. A large gap at high decode $\Rsq$ signals a quantity
represented but not used; a positive next-state $\tcs$ signals genuine deployment. The algebraic predicate
of \S\ref{sec:prelim} says in advance which invariants read as confounded (those affine in the output) and
should be scoped as a boundary. None of this requires retraining, and all of it composes with the
causal-abstraction machinery already standard in interpretability.

\noindent\textbf{Implications for the world-model debate.}
Debates over whether sequence and dynamics models build \emph{world models} often turn on probing
evidence: a quantity is decodable, therefore the model is said to represent the world's structure. Our
results show decodability is the wrong adjudicator, because presence and use dissociate lawfully, and two
models with identical decodable images may generalize completely differently. Whether a model has a
\emph{usable} world model is a question about its computation, which the deployment gap measures directly
and cheaply, well beyond physics.

\noindent\textbf{Limitations.}
The instrument reads deployment along a fitted linear direction; because the deployed invariant is
redundantly encoded, single-step interchange under-reads and our load-bearing values are conservative
lower bounds (\S\ref{sec:robust}). Linear/extensive invariants are entangled with linear output maps and
we scope them as a confounded boundary rather than a clean case. Our downstream payoff is demonstrated on
a controlled shortcut task; extending the deployment metric to a large model's end task is the natural
next step, for which the \Poseidon\ result is the first evidence at scale. We make claims only about the
systems and models tested and do not assert universality beyond them.

\newpage
\noindent\textbf{Reproducibility Statement.}
Every quantitative claim is drawn from a verified experiment log. Section~\ref{sec:prelim} states the
estimand exactly: decodability (Eq.~\ref{eq:decode}), the single-step interchange (Eq.~\ref{eq:interchange}),
and the transfer-corr readout (Eq.~\ref{eq:tc}). Section~\ref{sec:method} and Appendix~\ref{app:method}
specify the decode fit ($z$-scored cross-validated ridge), the donor-patch sampling, the two null controls,
and the readout heads. Appendix~\ref{app:results} lists, per system, the exact architectures, integrators
and step sizes, invariants, layer indices, seeds, decode $\Rsq$, interchange slopes, $\tcs$, ratios to the
random-direction null, and $95\%$ bootstrap confidence intervals with permutation $p$-values. The foundation-
model study (\S\ref{sec:scale}) uses the publicly released \Poseidon\ checkpoint \citep{herde2024poseidon}
with frozen weights and the layer indices and decode/interchange settings given in Appendix~\ref{app:scale}.
All interventions are single forward steps on frozen models; no training of $f_\theta$ is required to
reproduce any result.

\noindent\textbf{Ethics Statement.}
This work studies interpretability of models of physical dynamical systems (pendulums, circuits, and PDE
flows) and a pretrained scientific foundation model. It uses no human subjects, no personal or sensitive
data, and no generative content with foreseeable dual-use or safety concerns. All models are either
trained by us on synthetic physics simulations or are publicly released scientific checkpoints used under
their licenses. We see no negative societal impacts specific to this study; if anything, the central
message, that decodability overstates what a model uses, counsels caution before deploying probe-based
claims of model understanding in high-stakes scientific settings.

\bibliographystyle{plainnat}
\bibliography{references}

@article{iten2020scinet,
  title={Discovering physical concepts with neural networks},
  author={Iten, Raban and Metger, Tony and Wilming, Henrik and del Rio, L{\'\i}dia and Renner, Renato},
  journal={Physical Review Letters}, volume={124}, number={1}, pages={010508}, year={2020}, publisher={APS}}

@inproceedings{vafa2024worldmodel,
  title={Evaluating the world model implicit in a generative model},
  author={Vafa, Keyon and Chen, Justin Y. and Rambachan, Ashesh and Kleinberg, Jon and Mullainathan, Sendhil},
  booktitle={Advances in Neural Information Processing Systems (NeurIPS)}, year={2024}}

@inproceedings{vafa2025inductivebias,
  title={What has a foundation model found? {U}sing inductive bias to probe for world models},
  author={Vafa, Keyon and Chang, Peter G. and Rambachan, Ashesh and Mullainathan, Sendhil},
  booktitle={International Conference on Machine Learning (ICML)}, year={2025}}

@article{liu2021aipoincare,
  title={Machine learning conservation laws from trajectories},
  author={Liu, Ziming and Tegmark, Max},
  journal={Physical Review Letters}, volume={126}, number={18}, pages={180604}, year={2021}, publisher={APS}}

@article{ha2021conservnet,
  title={Discovering conservation laws from trajectories via machine learning},
  author={Ha, Seungwoong and Jeong, Hawoong},
  journal={arXiv preprint arXiv:2102.04008}, year={2021}}

@article{lemos2023orbital,
  title={Rediscovering orbital mechanics with machine learning},
  author={Lemos, Pablo and Jeffrey, Niall and Cranmer, Miles and Ho, Shirley and Battaglia, Peter},
  journal={Machine Learning: Science and Technology}, volume={4}, number={4}, pages={045002}, year={2023}, publisher={IOP Publishing}}

@inproceedings{greydanus2019hnn,
  title={Hamiltonian neural networks},
  author={Greydanus, Samuel and Dzamba, Misko and Yosinski, Jason},
  booktitle={Advances in Neural Information Processing Systems (NeurIPS)}, year={2019}}

@inproceedings{cranmer2020lnn,
  title={Lagrangian neural networks},
  author={Cranmer, Miles and Greydanus, Sam and Hoyer, Stephan and Battaglia, Peter and Spergel, David and Ho, Shirley},
  booktitle={ICLR Workshop on Integration of Deep Neural Models and Differential Equations}, year={2020}}

@inproceedings{alet2021noether,
  title={Noether networks: {M}eta-learning useful conserved quantities},
  author={Alet, Ferran and Doblar, Dylan and Zhou, Allan and Tenenbaum, Josh and Kawaguchi, Kenji and Finn, Chelsea},
  booktitle={Advances in Neural Information Processing Systems (NeurIPS)}, year={2021}}

@inproceedings{cranmer2020discovering,
  title={Discovering symbolic models from deep learning with inductive biases},
  author={Cranmer, Miles and Sanchez-Gonzalez, Alvaro and Battaglia, Peter and Xu, Rui and Cranmer, Kyle and Spergel, David and Ho, Shirley},
  booktitle={Advances in Neural Information Processing Systems (NeurIPS)}, year={2020}}

@article{udrescu2020aifeynman,
  title={{AI} {F}eynman: {A} physics-inspired method for symbolic regression},
  author={Udrescu, Silviu-Marian and Tegmark, Max},
  journal={Science Advances}, volume={6}, number={16}, pages={eaay2631}, year={2020}, publisher={AAAS}}

@inproceedings{herde2024poseidon,
  title={Poseidon: {E}fficient foundation models for {PDE}s},
  author={Herde, Maximilian and Raoni{\'c}, Bogdan and Rohner, Tobias and K{\"a}ppeli, Roger and Molinaro, Roberto and de B{\'e}zenac, Emmanuel and Mishra, Siddhartha},
  booktitle={Advances in Neural Information Processing Systems (NeurIPS)}, year={2024}}

@article{alain2017probes,
  title={Understanding intermediate layers using linear classifier probes},
  author={Alain, Guillaume and Bengio, Yoshua},
  journal={ICLR Workshop}, year={2017}}

@inproceedings{hewitt2019structural,
  title={A structural probe for finding syntax in word representations},
  author={Hewitt, John and Manning, Christopher D.},
  booktitle={Proceedings of NAACL-HLT}, year={2019}}

@inproceedings{hewitt2019control,
  title={Designing and interpreting probes with control tasks},
  author={Hewitt, John and Liang, Percy},
  booktitle={Proceedings of EMNLP-IJCNLP}, year={2019}}

@article{belinkov2022probing,
  title={Probing classifiers: {P}romises, shortcomings, and advances},
  author={Belinkov, Yonatan},
  journal={Computational Linguistics}, volume={48}, number={1}, pages={207--219}, year={2022}}

@inproceedings{li2023othello,
  title={Emergent world representations: {E}xploring a sequence model trained on a synthetic task},
  author={Li, Kenneth and Hopkins, Aspen K. and Bau, David and Vi{\'e}gas, Fernanda and Pfister, Hanspeter and Wattenberg, Martin},
  booktitle={International Conference on Learning Representations (ICLR)}, year={2023}}

@inproceedings{nanda2023othello,
  title={Emergent linear representations in world models of self-supervised sequence models},
  author={Nanda, Neel and Lee, Andrew and Wattenberg, Martin},
  booktitle={BlackboxNLP Workshop at EMNLP}, year={2023}}

@article{park2023linear,
  title={The linear representation hypothesis and the geometry of large language models},
  author={Park, Kiho and Choe, Yo Joong and Veitch, Victor},
  journal={arXiv preprint arXiv:2311.03658}, year={2023}}

@inproceedings{geiger2021causal,
  title={Causal abstractions of neural networks},
  author={Geiger, Atticus and Lu, Hanson and Icard, Thomas and Potts, Christopher},
  booktitle={Advances in Neural Information Processing Systems (NeurIPS)}, year={2021}}

@inproceedings{geiger2023das,
  title={Finding alignments between interpretable causal variables and distributed neural representations},
  author={Geiger, Atticus and Wu, Zhengxuan and Potts, Christopher and Icard, Thomas and Goodman, Noah D.},
  booktitle={Conference on Causal Learning and Reasoning (CLeaR)}, year={2024}}

@inproceedings{vig2020mediation,
  title={Investigating gender bias in language models using causal mediation analysis},
  author={Vig, Jesse and Gehrmann, Sebastian and Belinkov, Yonatan and Qian, Sharon and Nevo, Daniel and Singer, Yaron and Shieber, Stuart},
  booktitle={Advances in Neural Information Processing Systems (NeurIPS)}, year={2020}}

@inproceedings{meng2022rome,
  title={Locating and editing factual associations in {GPT}},
  author={Meng, Kevin and Bau, David and Andonian, Alex and Belinkov, Yonatan},
  booktitle={Advances in Neural Information Processing Systems (NeurIPS)}, year={2022}}

@inproceedings{wu2023boundless,
  title={Interpretability at scale: {I}dentifying causal mechanisms in {A}lpaca},
  author={Wu, Zhengxuan and Geiger, Atticus and Icard, Thomas and Potts, Christopher and Goodman, Noah D.},
  booktitle={Advances in Neural Information Processing Systems (NeurIPS)}, year={2023}}

@inproceedings{wang2023ioi,
  title={Interpretability in the wild: {A} circuit for indirect object identification in {GPT}-2 small},
  author={Wang, Kevin and Variengien, Alexandre and Conmy, Arthur and Shlegeris, Buck and Steinhardt, Jacob},
  booktitle={International Conference on Learning Representations (ICLR)}, year={2023}}

@article{bricken2023monosemanticity,
  title={Towards monosemanticity: {D}ecomposing language models with dictionary learning},
  author={Bricken, Trenton and Templeton, Adly and Batson, Joshua and others},
  journal={Transformer Circuits Thread}, year={2023}}

@inproceedings{cunningham2023sparse,
  title={Sparse autoencoders find highly interpretable features in language models},
  author={Cunningham, Hoagy and Ewart, Aidan and Riggs, Logan and Huben, Robert and Sharkey, Lee},
  booktitle={International Conference on Learning Representations (ICLR)}, year={2024}}

@article{geirhos2020shortcut,
  title={Shortcut learning in deep neural networks},
  author={Geirhos, Robert and Jacobsen, J{\"o}rn-Henrik and Michaelis, Claudio and Zemel, Richard and Brendel, Wieland and Bethge, Matthias and Wichmann, Felix A.},
  journal={Nature Machine Intelligence}, volume={2}, number={11}, pages={665--673}, year={2020}}

@inproceedings{elhage2022superposition,
  title={Toy models of superposition},
  author={Elhage, Nelson and Hume, Tristan and Olsson, Catherine and others},
  booktitle={Transformer Circuits Thread}, year={2022}}

@inproceedings{conneau2018probing,
  title={What you can cram into a single vector: {P}robing sentence embeddings for linguistic properties},
  author={Conneau, Alexis and Kruszewski, German and Lample, Guillaume and Barrault, Lo{\"i}c and Baroni, Marco},
  booktitle={Proceedings of the 56th Annual Meeting of the Association for Computational Linguistics (ACL)}, year={2018}}

@inproceedings{tenney2019bert,
  title={{BERT} rediscovers the classical {NLP} pipeline},
  author={Tenney, Ian and Das, Dipanjan and Pavlick, Ellie},
  booktitle={Proceedings of the 57th Annual Meeting of the Association for Computational Linguistics (ACL)}, year={2019}}

@inproceedings{conmy2023acdc,
  title={Towards automated circuit discovery for mechanistic interpretability},
  author={Conmy, Arthur and Mavor-Parker, Augustine N. and Lynch, Aengus and Heimersheim, Stefan and Garriga-Alonso, Adri{\`a}},
  booktitle={Advances in Neural Information Processing Systems (NeurIPS)}, year={2023}}

@inproceedings{nanda2023grokking,
  title={Progress measures for grokking via mechanistic interpretability},
  author={Nanda, Neel and Chan, Lawrence and Lieberum, Tom and Smith, Jess and Steinhardt, Jacob},
  booktitle={International Conference on Learning Representations (ICLR)}, year={2023}}

@article{olah2020circuits,
  title={Zoom in: {A}n introduction to circuits},
  author={Olah, Chris and Cammarata, Nick and Schubert, Ludwig and Goh, Gabriel and Petrov, Michael and Carter, Shan},
  journal={Distill}, year={2020}}

@inproceedings{gurnee2024space,
  title={Language models represent space and time},
  author={Gurnee, Wes and Tegmark, Max},
  booktitle={International Conference on Learning Representations (ICLR)}, year={2024}}

@article{templeton2024scaling,
  title={Scaling monosemanticity: {E}xtracting interpretable features from {C}laude 3 {S}onnet},
  author={Templeton, Adly and Conerly, Tom and Marcus, Jonathan and Lindsey, Jack and Bricken, Trenton and Chen, Brian and others},
  journal={Transformer Circuits Thread}, year={2024}}

@article{gao2024scaling,
  title={Scaling and evaluating sparse autoencoders},
  author={Gao, Leo and la Tour, Tom Dupr{\'e} and Tillman, Henk and Goh, Gabriel and Troll, Rajan and Radford, Alec and Sutskever, Ilya and Leike, Jan and Wu, Jeffrey},
  journal={arXiv preprint arXiv:2406.04093}, year={2024}}

@article{raissi2019pinn,
  title={Physics-informed neural networks: {A} deep learning framework for solving forward and inverse problems involving nonlinear partial differential equations},
  author={Raissi, Maziar and Perdikaris, Paris and Karniadakis, George E.},
  journal={Journal of Computational Physics}, volume={378}, pages={686--707}, year={2019}, publisher={Elsevier}}

@inproceedings{li2021fno,
  title={Fourier neural operator for parametric partial differential equations},
  author={Li, Zongyi and Kovachki, Nikola and Azizzadenesheli, Kamyar and Liu, Burigede and Bhattacharya, Kaushik and Stuart, Andrew and Anandkumar, Anima},
  booktitle={International Conference on Learning Representations (ICLR)}, year={2021}}

@article{lu2021deeponet,
  title={Learning nonlinear operators via {DeepONet} based on the universal approximation theorem of operators},
  author={Lu, Lu and Jin, Pengzhan and Pang, Guofei and Zhang, Zhongqiang and Karniadakis, George Em},
  journal={Nature Machine Intelligence}, volume={3}, number={3}, pages={218--229}, year={2021}, publisher={Nature Publishing Group}}

@inproceedings{sanchez2020simulate,
  title={Learning to simulate complex physics with graph networks},
  author={Sanchez-Gonzalez, Alvaro and Godwin, Jonathan and Pfaff, Tobias and Ying, Rex and Leskovec, Jure and Battaglia, Peter W.},
  booktitle={International Conference on Machine Learning (ICML)}, year={2020}}

@inproceedings{brandstetter2022mppde,
  title={Message passing neural {PDE} solvers},
  author={Brandstetter, Johannes and Worrall, Daniel E. and Welling, Max},
  booktitle={International Conference on Learning Representations (ICLR)}, year={2022}}

@inproceedings{battaglia2016interaction,
  title={Interaction networks for learning about objects, relations and physics},
  author={Battaglia, Peter W. and Pascanu, Razvan and Lai, Matthew and Rezende, Danilo Jimenez and Kavukcuoglu, Koray},
  booktitle={Advances in Neural Information Processing Systems (NeurIPS)}, year={2016}}

@article{brunton2016sindy,
  title={Discovering governing equations from data by sparse identification of nonlinear dynamical systems},
  author={Brunton, Steven L. and Proctor, Joshua L. and Kutz, J. Nathan},
  journal={Proceedings of the National Academy of Sciences}, volume={113}, number={15}, pages={3932--3937}, year={2016}}

@inproceedings{mikolov2013word2vec,
  title={Distributed representations of words and phrases and their compositionality},
  author={Mikolov, Tomas and Sutskever, Ilya and Chen, Kai and Corrado, Greg S. and Dean, Jeff},
  booktitle={Advances in Neural Information Processing Systems (NeurIPS)}, year={2013}}

@inproceedings{toshniwal2022chess,
  title={Chess as a testbed for language model state tracking},
  author={Toshniwal, Shubham and Wiseman, Sam and Livescu, Karen and Gimpel, Kevin},
  booktitle={Proceedings of the AAAI Conference on Artificial Intelligence}, year={2022}}

@inproceedings{jin2024program,
  title={Emergent representations of program semantics in language models trained on programs},
  author={Jin, Charles and Rinard, Martin},
  booktitle={International Conference on Machine Learning (ICML)}, year={2024}}

@article{raukert2023survey,
  title={Toward transparency in {AI}: {S}urvey on interpreting the inner structures of deep neural networks},
  author={R{\"a}uker, Tilman and Ho, Anson and Casper, Stephen and Hadfield-Menell, Dylan},
  journal={IEEE Conference on Secure and Trustworthy Machine Learning (SaTML)}, year={2023}}

\appendix
\newpage
\section{Instrument details}\label{app:method}
\noindent\textbf{Decode fit.} Unless noted, $w_\ell$ is fit by ridge regression on $z$-scored activations with
$5$-fold cross-validation; we report held-out $\Rsq$ and take $\uhat=w_\ell/\lVert w_\ell\rVert$. The
layer $\ell$ is chosen as the decode-maximizing hidden layer.
\noindent\textbf{Interchange sampling.} For each system we sample target/donor pairs $(s,s')$ i.i.d.\ from the
trajectory distribution, apply Eq.~\eqref{eq:interchange}, and collect $(\dphi,\dg)$ over the pair set;
$\beta$ is the OLS slope and $\tcs$ the Pearson correlation (Eq.~\ref{eq:tc}).
\noindent\textbf{Controls.} The shuffled-label probe refits $w_\ell$ on a random permutation of $\phivar$; the
matched random-direction control replaces $\uhat$ with a uniformly random unit vector rescaled to
$\lVert w_\ell\rVert$. Both are recomputed per seed and must read $\approx0$.

\section{System specifications and conserved-quantity definitions}\label{app:systems}
Table~\ref{tab:systems} lists every dynamical system in the study with its state dimension, invariant
class, and the closed-form conserved quantity used as ground truth, the exact answer key for both the
decode fit (Eq.~\ref{eq:decode}) and the invariant readout. All ODE systems are integrated with a
fixed-step RK4 scheme at $\Delta t\!=\!0.02$ (measured energy drift $6.5\!\times\!10^{-6}$ over the
horizon, far under the $10^{-2}$ abort threshold), and evaluation sets are hashed before analysis. Unless
noted, mechanical constants are normalized to unit mass, length, and coupling.

\begin{table}[h]
\centering\footnotesize
\setlength{\tabcolsep}{5pt}\renewcommand{\arraystretch}{1.3}
\caption{\textbf{Dynamical systems and their conserved quantities.} Nonlinear invariants (energy, angular
momentum) yield the clean decode--deploy dissociation; linear/extensive invariants (linear momentum) form
the confounded boundary of \S\ref{sec:taxonomy}.}
\label{tab:systems}
\begin{tabular}{@{}lcll@{}}
\toprule
System & Dim & Class & Conserved quantity $\phivar$ \\
\midrule
simple pendulum & $2$ & energy (nonlinear) & $H=\tfrac12\omega^{2}-g\cos\theta$ \\
LC circuit & $2$ & energy (nonlinear) & $E=\tfrac12 Li^{2}+\tfrac{1}{2C}q^{2}$ \\
two-body spring/collision & $4$ & linear mom.\ (linear) & $P=\textstyle\sum_i m_i v_i$ \\
$2$D central force & $4$ & angular mom.\ (bilinear) & $L=x\,v_y-y\,v_x$ \\
$1$D wave equation & $32$ & energy (nonlinear) & $E=\textstyle\sum_i\!\big(\tfrac12 w_i^{2}+\tfrac12 c^{2}(\partial_x u)_i^{2}\big)$ \\
Kepler two-body ($1/r^{2}$) & $4$ & energy; ang.\ mom. & $E=\tfrac12\lVert v\rVert^{2}-\tfrac1r$;\ \ $L=x v_y-y v_x$ \\
\bottomrule
\end{tabular}
\end{table}

\noindent\textbf{Dynamics.} The pendulum obeys $\dot\theta=\omega,\ \dot\omega=-g\sin\theta$; the LC circuit
$\dot q=i,\ \dot\imath=-q/(LC)$; the central-force particle moves under $\ddot{\mathbf r}=-\mathbf
r/\lVert\mathbf r\rVert^{3}$; the wave field integrates $u_{tt}=c^{2}u_{xx}$ with periodic boundaries on a
$32$-point grid, written as a first-order system in $(u,w{=}u_t)\in\R^{32}$. Kepler follows the same
inverse-square law as the central force but is evaluated in the chaotic, only-partially-decodable regime
($\Rsq_{\text{energy}}\!=\!0.80$) used as a calibration against prior world-model probes.

\noindent\textbf{Why nonlinear vs.\ linear invariants behave differently.} A linear/extensive invariant such as
$P=\sum_i v_i$ is affine in the velocity output coordinates, so its decode direction is entangled with the
linear output map and interchanging it mechanically moves the prediction, the confound of
\S\ref{sec:taxonomy} ($\tcs\!\approx\!-0.7$). A nonlinear invariant (energy $\tfrac12\omega^2-g\cos\theta$,
angular momentum $x v_y-y v_x$) has a vanishing linear part in the output on-distribution, so next-state
interchange is inert. The $\operatorname{asinh}(P)$ construction breaks the affine relation for the
\emph{same} $P$, flipping it from confounded to load-bearing.

\noindent\textbf{Invariants probed in the algebraic taxonomy.} The taxonomy of \S\ref{sec:taxonomy} contrasts
several invariants on the \emph{same} systems, so we list them explicitly. On the two-body spring:
\emph{linear momentum} $P=\sum_i v_i$ (affine in the velocity output $\Rightarrow$ strong confound,
$\tcs\!=\!-1.02$), \emph{spring energy} $E=\tfrac12\sum_i v_i^{2}+\tfrac12(x_1-x_2)^{2}$ (quadratic in the
outputs $\Rightarrow$ weak confound, $\tcs\!=\!-0.05$), and the same $P$ read through an
$\operatorname{asinh}(P)$ head (nonlinear output $\Rightarrow$ load-bearing, $\tcs\!=\!+0.60$). On the wave
field: \emph{energy} $E=\sum_i\!\big(\tfrac12 w_i^{2}+\tfrac12 c^{2}(\partial_x u)_i^{2}\big)$ (nonlinear
$\Rightarrow$ clean dissociation) versus \emph{total momentum} $P=\sum_i w_i$ (linear $\Rightarrow$
confound, $\tcs\!=\!-0.71$). Every point on the overlap spectrum of Figure~\ref{fig:results}c is one of
these closed-form invariants; the verdict tracks the invariant's algebraic relation to the output, not its
physical identity.

\noindent\textbf{Readout targets.} The next-state readout $g_{\text{next}}$ is the model's own one-step
prediction projected onto the invariant-sensitive output coordinate. The invariant readout
$g_{\text{inv}}$ is a small head trained to output $\phivar$ from the frozen features (held-out fit
$\Rsq\!\ge\!0.99$). For the taxonomy flip we additionally train an $\operatorname{asinh}(P)$ head on the
\emph{same} linear-momentum system, changing only the algebra of the output while holding the invariant
and the representation fixed.

\noindent\textbf{Architectures and training.} Each system is learned by three vanilla backbones with no
conservation prior and a shared parameter budget ($<\!10^{5}$ parameters): an MLP (default), a GRU, and a
short-context Transformer; the recurrent and attention models read a history window rather than a single
state, so inertness there cannot be a feed-forward artifact. Networks are trained to convergence on the
one-step prediction (or the invariant target) and then frozen; every intervention is on frozen weights.
The capacity sweep of \S\ref{sec:robust} varies the MLP hidden width over $\{4,8,16,32,64\}$. In total the
vanilla matrix comprises on the order of a hundred such networks, each under $10^{5}$ parameters and
trainable in seconds to minutes on CPU, so the coverage across substrates, dimensions, invariant classes,
and architectures is obtained at near-zero compute.

\section{Sparse-autoencoder localization of the deployed invariant}\label{app:sae}
To test whether the deployed invariant occupies an identifiable, sparse set of features rather than being
smeared across the representation, we fit a small overcomplete sparse autoencoder (SAE) on the frozen
activations of the pendulum invariant-target network (one hidden layer, width $16$; SAE $16\!\to\!64$, ReLU with
an $L_1$ penalty; reconstruction $\Rsq\!=\!1.000$, $58$ active latents). The latent best aligned with the
invariant reaches $\lvert\corr\rvert\!=\!0.84$, and ablating it degrades the head's invariant prediction
$7\times$ more than ablating a random latent (drift $0.015$ vs.\ $0.0022$; the random-latent control is
$\approx0$). The small absolute drift reflects the same redundancy seen elsewhere, the invariant is
spread over the $\sim\!58$ active latents, so single-latent ablation under-reads
(Appendix~\ref{app:robust}). Crucially, the load-bearing direction is \emph{found} in an unsupervised SAE
basis rather than built into a designed readout, distinguishing the diagnosis from methods that engineer
the invariant in by construction.

\section{Per-system results, statistics, and boundary}\label{app:results}
Table~\ref{tab:main} in the main text gives the headline matrix; Table~\ref{tab:persystem} records the full
per-system readings (visualized in Figure~\ref{fig:decouple}), and below we give the accompanying bootstrap statistics.

\begin{table}[h]
\centering\footnotesize
\setlength{\tabcolsep}{3.5pt}\renewcommand{\arraystretch}{1.2}
\caption{\textbf{Full per-system transfer-corr results} (primary single-direction readout, mean over
seeds). Nonlinear invariants are \present{inert on next-state} and \deployed{load-bearing} under the
invariant objective; linear-momentum rows are the \confound{confounded boundary}; the final row is the
$\operatorname{asinh}(P)$ flip of \S\ref{sec:taxonomy}. \emph{ratio} is the load-bearing slope over the
matched random-direction null. Self-repair index $0.76$--$0.999$ across the pendulum, LC, central-force, and wave systems.}
\label{tab:persystem}
\begin{tabular}{@{}llccccc@{}}
\toprule
System & Invariant (class) & decode $\Rsq$ & next-state $\tcs$ & inv-obj $\tcs$ & ratio & seeds \\
\midrule
pendulum (MLP)      & energy (nonlinear)        & $1.00$ & \present{$-0.01$} & \deployed{$+0.97$} & $25.1\times$ & $10$ \\
pendulum (GRU)      & energy (nonlinear)        & $0.98$ & \present{$+0.02$} & \deployed{$+0.79$} & $-$ & $5$ \\
pendulum (Transf.)  & energy (nonlinear)        & $0.98$ & \present{$+0.02$} & \deployed{$+0.70$} & $-$ & $5$ \\
LC circuit          & energy (nonlinear)        & $0.99$ & \present{$+0.00$} & \deployed{$+0.31$} & $41\times$ & $5$ \\
central force       & angular mom.\ (bilinear)  & $0.99$ & \present{$+0.02$} & \deployed{$+1.00$} & $37\times$ & $5$ \\
wave PDE ($32$-d)   & energy (nonlinear)        & $0.99$ & \present{$+0.00$} & \deployed{$+0.97$} & $175\times$ & $5$ \\
Kepler $1/r^{2}$       & energy (entangled)        & $0.80$ & \present{$-0.02$} & \deployed{$+0.39$} & $40\times$ & $5$ \\
Kepler $1/r^{2}$       & angular momentum          & $0.99$ & \present{$+0.04$} & \deployed{$+0.35$} & $10\times$ & $5$ \\
\midrule
\rowcolor{CONFOUNDbg} spring & linear mom.\ (linear)  & $0.98$ & \confound{$-1.02$} & $-$ & $-$ & $5$ \\
\rowcolor{CONFOUNDbg} wave   & linear mom.\ (linear)  & $0.99$ & \confound{$-0.71$} & $-$ & $-$ & $5$ \\
\rowcolor{DEPLOYbg} spring, $\operatorname{asinh}(P)$ & linear mom.\ (nonlin.\ out) & $0.98$ & $-$ & \deployed{$+0.60$} & $-$ & $3$ \\
\bottomrule
\end{tabular}
\end{table}

The bootstrap/permutation pass (conservative single-direction estimator) yields, for
the pendulum and central-force anchors: pendulum next-state $\tcs\!=\!-0.010$, CI $[-0.019,+0.001]$,
$p\!\approx\!0.73$ (n.s.); pendulum deployment $\tcs\!=\!+0.173$, CI $[+0.061,+0.284]$, $p\!=\!0.0033$;
central-force next-state $\tcs\!=\!-0.012$, CI $[-0.023,-0.001]$, $p\!\approx\!0.71$ (n.s.); central-force
deployment $\tcs\!=\!+0.733$, CI $[+0.667,+0.794]$, $p\!=\!0.0033$. The conservative estimator gives smaller
magnitudes than the primary readout used in Table~\ref{tab:main}; the sign and significance structure
(inert indistinguishable from zero, deployment significantly positive) is what the statistics establish.

\begin{table}[h]
\centering\footnotesize
\setlength{\tabcolsep}{5pt}\renewcommand{\arraystretch}{1.25}
\caption{\textbf{Bootstrap ($95\%$ CI) and permutation ($300$-shuffle) statistics} on the representative
anchors ($10$ seeds, conservative single-direction estimator). Inert conditions are
indistinguishable from zero; deployment conditions exclude zero with $p\!=\!0.0033$.}
\label{tab:stats}
\begin{tabular}{@{}llccc@{}}
\toprule
System & Condition & $\tcs$ (95\% CI) & slope $\beta$ (95\% CI) & perm.\ $p$ \\
\midrule
pendulum      & next-state (\present{inert})    & $-0.010\ [-0.019,+0.001]$ & $-$ & $0.73$ (n.s.) \\
pendulum      & deployment (\deployed{LB})      & $+0.173\ [+0.061,+0.284]$ & $+0.150\ [+0.058,+0.238]$ & $\mathbf{0.0033}$ \\
central force & next-state (\present{inert})    & $-0.012\ [-0.023,-0.001]$ & $-$ & $0.71$ (n.s.) \\
central force & deployment (\deployed{LB})      & $+0.733\ [+0.667,+0.794]$ & $+0.554\ [+0.498,+0.613]$ & $\mathbf{0.0033}$ \\
\bottomrule
\end{tabular}
\end{table}

\label{app:robust}
\begin{figure}[t]
\centering
\includegraphics[width=\textwidth]{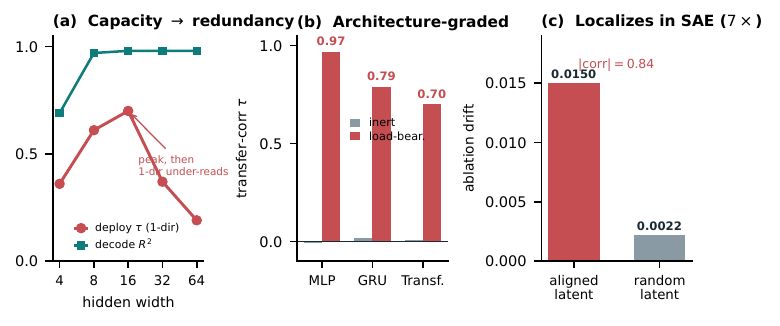}
\caption{\textbf{The deployed invariant is redundantly encoded (so our numbers are lower bounds).}
\textbf{(a)}~Deployment strength is non-monotonic in width, peaking at intermediate capacity, evidence of
distributed, redundant coding rather than a single privileged unit. \textbf{(b)}~Inertness on next-state
is architecture-agnostic; deployment magnitude is graded. \textbf{(c)}~A sparse-autoencoder latent aligns
with the invariant at $|\text{corr}|\!=\!0.84$ and its ablation exceeds the random-direction null by
$7\times$, localizing the deployed direction.}
\label{fig:redundancy}
\end{figure}
\noindent\textbf{Redundancy and capacity.} The self-repair index (fraction of the interchange effect restored by
the rest of the network within one step) ranges $0.76$--$0.999$ across the pendulum, LC, central-force, and wave systems. Deployment strength is
non-monotonic in hidden width while decodability saturates early (Table~\ref{tab:capacity},
Figure~\ref{fig:redundancy}a): as capacity grows the invariant is coded more redundantly, so
single-direction interchange progressively under-reads it. A sparse-autoencoder latent aligns with the
invariant at $|\text{corr}|\!=\!0.84$ and its ablation drives a drift of $0.015$ versus $0.0022$ for a
random direction ($7\times$).

\begin{table}[h]
\centering\footnotesize
\setlength{\tabcolsep}{8pt}\renewcommand{\arraystretch}{1.2}
\caption{\textbf{Capacity sweep (pendulum, invariant-target head).} Decode $\Rsq$ saturates by width
$8$; single-direction deployment $\tcs$ peaks at width $16$ and then declines as redundancy grows and the
one-direction probe under-reads, not a weakening of deployment but a widening lower-bound gap.}
\label{tab:capacity}
\begin{tabular}{@{}lccccc@{}}
\toprule
hidden width & $4$ & $8$ & $16$ & $32$ & $64$ \\
\midrule
decode $\Rsq$ & $0.69$ & $0.97$ & $0.98$ & $0.98$ & $0.98$ \\
deployment $\tcs$ (1-dir) & $+0.36$ & $+0.61$ & $\mathbf{+0.70}$ & $+0.37$ & $+0.19$ \\
\bottomrule
\end{tabular}
\end{table}
\noindent\textbf{Linear boundary.} Spring/wave linear momentum reads $\tcs\!=\!-0.69/-0.71$ on next-state: the
decoded direction is entangled with the linear output map, so the interchange moves the output for an
algebraic reason. We report these as a confounded boundary (shaded, Table~\ref{tab:main}); the
$\operatorname{asinh}(P)$ flip of \S\ref{sec:taxonomy} confirms the entanglement is algebraic, not
physical.

\noindent\textbf{Additional controls and observations.} Three further checks support the reading.
\emph{(i)~Inertness is not a bottleneck or probe-dimension artifact.} On the pendulum, next-state
inertness holds across layers $L_1/L_2$ and across $\{1,5,16\}$-dimensional energy subspaces (all slopes
$\approx$ random), and a deliberately tight two-dimensional bottleneck stores the \emph{state} rather than
energy (energy decode $\Rsq\!\approx\!0.003$ there), so the causal inertness cannot be blamed on a probe
that merely missed the invariant. \emph{(ii)~The model need not even represent what it does not use.} On
the $32$-dimensional wave PDE the next-state network decodes energy at only $\Rsq\!=\!0.13$, it never
builds the quadratic energy feature it does not need, while the invariant-target network on the same
substrate decodes it at $0.99$; presence itself, not only deployment, follows the objective. \emph{(iii)~Why
the dissociation is so clean.} Next-state prediction on the full microstate is the \emph{worst-case}
objective for finding a load-bearing invariant: the microstate is a sufficient statistic for the next
state, so any conserved quantity is a redundant function of it and never a needed causal coordinate. An
invariant-relevant objective removes that redundancy, which is precisely when the same decoded direction
becomes load-bearing.

\section{Foundation-model study details}\label{app:scale}
\noindent\textbf{Checkpoint and hooks.} We use the public \Poseidon\ checkpoint ($158$M parameters, an scOT
SwinV2 U-Net) with frozen weights. The bottleneck probe hooks the deepest encoder level
(\texttt{encoder.layers.3.blocks.7}, feature dimension $768$); the output-adjacent probe hooks a decoder
layer separated from the field prediction by the remaining upsampling stages and the projection head.
Interventions use the same single-step interchange (Algorithm~\ref{alg:probe}) as the vanilla matrix, at
lead time $t\!=\!1.0$ with $N\!=\!512$ pairs.

\noindent\textbf{Two conserved quantities.} The nonlinear invariant is kinetic energy
$\mathrm{KE}=\int\tfrac12(u^2+v^2)$ (approximately conserved for incompressible Navier--Stokes); the linear
functional is $\int u$. For the depth study, inputs are divergence-free velocity fields (spectral
projection, divergence RMS $\approx\!5\!\times\!10^{-7}$) assembled to the four-channel $[\rho,u,v,p]$
layout and normalized with the exact scOT constants, in-distribution valid inputs, not simulation
trajectories. The headline replication additionally uses $512$ real NS-Sines trajectories, which removes
any synthetic-input assumption.

\noindent\textbf{Results.} On real trajectories, kinetic energy is decodable at $\Rsq\!=\!0.916$ yet inert on
the model's own one-step prediction ($\tcs\!=\!+0.06$), while an invariant readout on the \emph{same}
frozen bottleneck features is load-bearing (readout fit $\Rsq\!=\!0.999$, slope $+0.616$,
$\tcs\!=\!+0.85$), both halves of the dissociation at scale. Table~\ref{tab:fm} gives the depth-resolved
interchange: at the encoder bottleneck both invariants are inert, whereas at the output-adjacent decoder
layer the linear $\int u$ becomes entangled while nonlinear KE stays inert, with the linear slope
$\sim\!800\times$ the nonlinear one at the same layer.

\begin{table}[h]
\centering\footnotesize
\setlength{\tabcolsep}{6pt}\renewcommand{\arraystretch}{1.25}
\caption{\textbf{Depth-resolved interchange on Poseidon-B~\citep{herde2024poseidon} (158M, in-distribution inputs).} Both invariants
are inert at the bottleneck; only the \confound{linear} functional entangles at the output-adjacent
decoder layer, exactly the taxonomy of \S\ref{sec:taxonomy} resolved along depth.}
\label{tab:fm}
\begin{tabular}{@{}llcccl@{}}
\toprule
Layer & Invariant (class) & decode $\Rsq$ & slope $\beta$ & $\tcs$ & verdict \\
\midrule
encoder bottleneck        & KE (nonlinear)   & $0.850$ & $+0.0032$ & \present{$+0.06$} & inert \\
encoder bottleneck        & $\int u$ (linear)& $0.998$ & $-0.0007$ & \present{$+0.06$} & inert \\
decoder (output-adjacent) & KE (nonlinear)   & $0.936$ & $+0.0002$ & \present{$-0.04$} & inert \\
decoder (output-adjacent) & $\int u$ (linear)& $1.000$ & $+0.1582$ & \confound{$+0.36$} & entangled \\
\bottomrule
\end{tabular}
\end{table}

\noindent\textbf{Depth resolves an apparent contradiction.} At the bottleneck the linear invariant is
\emph{inert}, contrary to the naive taxonomy prediction, because the bottleneck is separated from the
output by the entire nonlinear decoder, which breaks the affine invariant--output relation; on a shallow
residual-CNN stand-in the same $\int u$ does entangle (slope $+0.117$). Deployment on next-state prediction
therefore depends not only on the invariant--output algebra (\S\ref{sec:taxonomy}) but on the probe's
distance from the output. \emph{Honest null caveat:} the linear random-direction baseline is nonzero
($+0.022$), the same $k\!>\!1$ leakage of redundant/linear invariants into random subspaces noted in
\S\ref{sec:method}; the load-bearing evidence at the decoder is the directed slope ($7\times$ random) and
the same-layer nonlinear contrast ($800\times$), not a ratio-to-random, and all values are subject to the
single-direction lower bound.

\section{Extended related work and differentiation}\label{app:related}
Table~\ref{tab:related} places our contribution against the closest lines of work; here we differentiate
in more detail along the axes that matter.

\noindent\textbf{Designed readouts and physics-prior networks.} SciNet recovers physical parameters from a
\emph{designed} bottleneck \citep{iten2020scinet}, and conservation-prior architectures build the invariant
in by construction \citep{greydanus2019hnn,cranmer2020lnn,alet2021noether}. Both establish that an
invariant \emph{can be} represented, by engineering it in or reading it out of a hand-chosen layer. We
neither engineer nor hand-place: the deployed direction is \emph{found} in an unsupervised
sparse-autoencoder basis (Appendix~\ref{app:sae}), and the object of study is not presence but whether the
frozen computation causally \emph{uses} it, and under what objective.

\noindent\textbf{Symbolic law discovery.} Pipelines that distill governing equations from data
\citep{cranmer2020discovering,udrescu2020aifeynman,brunton2016sindy,liu2021aipoincare,ha2021conservnet} answer whether a law is
\emph{recoverable} from a trained model or its trajectories. This is a presence/recoverability question
under a single objective; it says nothing about whether the model's own forward computation depends on the
recovered quantity, which is exactly the gap we measure.

\noindent\textbf{World-model probing on dynamics models.} Recent work probes the inductive biases of
sequence and dynamics models on physical systems and asks whether prediction-trained models internalize the
right invariants \citep{vafa2024worldmodel,vafa2025inductivebias,lemos2023orbital}. These are
presence-centric analyses on a fixed objective. Our contribution is orthogonal and complementary: the
\emph{objective- and algebra-dependence} of causal deployment for a continuous conserved quantity, and the
demonstration that the deployment gap, not decodability, forecasts out-of-distribution behavior. We use
Kepler only as a calibration point against this line, not as a headline, and make no phase-transition or
capacity-threshold claim.

\noindent\textbf{Emergent-world-model probe-and-intervene.} Studies that combine probing with intervention to
argue a model \emph{uses} a learned latent \citep{li2023othello,nanda2023othello} are the closest in
method. Our instrument is in the same causal-abstraction family
\citep{geiger2021causal,geiger2023das,wu2023boundless,meng2022rome,vig2020mediation}, and we explicitly do
\emph{not} claim it as a contribution. What is new is the target of the intervention, a continuous
physical invariant, and the questions asked of it: \emph{when} deployment occurs (a testable algebraic
predicate), that it is set by the training objective rather than the representation, and that the resulting
gap is predictive of generalization.

\noindent\textbf{Superposition and shortcuts.} That a decodable feature can be causally inert is consistent with
distributed, redundant coding \citep{elhage2022superposition,bricken2023monosemanticity,cunningham2023sparse}
and with shortcut learning \citep{geirhos2020shortcut}. We make this precise for physical invariants,
quantify the redundancy that turns single-direction interchange into a lower bound (Appendix~\ref{app:sae},
Table~\ref{tab:capacity}), and connect the shortcut/robust split to the deployment gap
(\S\ref{sec:payoff}).

\paragraph{Neural dynamics surrogates.} The models we probe are themselves learned surrogates for
dynamical systems, and our question is orthogonal to their predictive accuracy. The substrate spans
physics-informed networks \citep{raissi2019pinn}, neural operators \citep{li2021fno,lu2021deeponet},
graph-network and interaction-network simulators \citep{battaglia2016interaction,sanchez2020simulate,brandstetter2022mppde},
and PDE foundation models \citep{herde2024poseidon}; for any such trained surrogate we ask only whether its
forward computation \emph{uses} a quantity it demonstrably represents, independent of how accurately it
rolls the dynamics forward.

\end{document}